\documentclass[sigconf,natbib=true,nonacm]{acmart}

\setcopyright{none}
\acmYear{2026}
\acmISBN{}
\acmDOI{}

\usepackage{cleveref}
\usepackage{enumitem}  
\usepackage{tabularx}
\usepackage{arydshln}
\usepackage{multirow}
\usepackage{float}
\usepackage{placeins}  
\usepackage[htt]{hyphenat}

\usepackage{listings}
\definecolor{codebg}{gray}{0.96}
\title{Zero-Trust Authorization and Discovery for Enterprise MCP}

\author{Huan Li}
\affiliation{\institution{PayPal Inc.}\country{USA}}
\email{huanli1@paypal.com}
\author{Yuwei Wang}
\affiliation{\institution{PayPal Inc.}\country{USA}}
\email{yuwewang@paypal.com}
\author{Srinivasan Manoharan}
\affiliation{\institution{PayPal Inc.}\country{USA}}
\email{srinivmanoharan@paypal.com}

\begin{document}

\begin{abstract}
LLM agents translate natural-language context, which may include attacker-controlled text, into privileged tool calls, so authorization must remain effective even when an agent is prompt-injected or otherwise adversarially steered. The Model Context Protocol (MCP) has become a widely adopted interface for this boundary, yet the authentication and authorization primitives in its official SDKs fall short of enterprise zero-trust requirements---most acutely the \emph{dual-persona} model, in which one server must serve human users (corporate SSO) and automated agents (service-account credentials in a different header) at once. We conduct a systematic gap analysis of six surveyed MCP SDKs (Python, TypeScript, Go, Rust, C\#, Swift) and identify three structural shortcomings: Authorization-header-bound credential extraction that makes dual-persona deployment difficult without replacing SDK-level middleware, the absence of pre-authentication tool discovery, and the lack of fine-grained per-tool authorization in the base SDKs. We close these gaps with composable extensions to FastMCP---cross-header credential normalization for enterprise deployments serving both human and service-account callers, cached token verification across heterogeneous IdPs, an unauthenticated metadata endpoint for credential-free registry discovery, and permission-filtered tool visibility kept consistent with per-tool invocation enforcement by a single declarative annotation---all implemented without modifying the protocol or SDK internals. Across four frontier LLMs over 2160 attempts, an in-body-check-only server still exposes forbidden tools ($152/720$, $21.1\%$; per-model rates $0\%$ to $50\%$ on identical prompts), whereas permission-aware visibility drives the rate to $0/720$; visibility-only filtering remained bypassable by scripted clients, while models continued to reference the hidden tool by name in up to 94\% of settings where the injected prompt let the model infer that name, confirming that discovery controls cannot replace invocation-time enforcement.

\end{abstract}


\maketitle

\section{Introduction}
\label{sec:introduction}

LLM agents introduce an authorization adversary that traditional API security was not designed to contain. Their privileged tool calls are generated from natural-language context that may include untrusted text: retrieved documents, third-party tool outputs, user prompts, or attacker-controlled web content. The agent thus becomes a vehicle for indirect prompt injection~\cite{greshake2023indirect,zhan2024injecagent}: network-layer authentication can verify the credential on a request, but not whether malicious context shaped the requested action. The attacker is not at the transport boundary; the attacker is \emph{inside the caller's reasoning loop}. Authorization must therefore be enforced at the MCP tool-invocation layer, before model-generated requests reach concrete privileged operations.

The Model Context Protocol (MCP)~\cite{anthropic2024mcp}, introduced in late 2024, has become a widely adopted interface where this enforcement boundary lives. MCP standardizes how LLM agents discover and invoke external tools, prompts, and resources. Within just over a year of release, enterprises are deploying dozens to hundreds of MCP servers exposing privileged operations: querying production databases, triggering CI/CD pipelines, managing Kubernetes clusters, and reading proprietary codebases. Each registered tool is a new attack surface: an agent with deployment access can deploy code; one with database access can exfiltrate data. Unlike a human API caller, the agent may attempt these operations because an attacker-controlled string in its context window told it to. This risk is catalogued in the 2025 OWASP LLM Top-10 as ``Excessive Agency''~\cite{owasp2025llmtop10} and flagged as an emerging surface in the NIST AI Risk Management Framework~\cite{nist2023airmf}.

\textbf{This work positions zero-trust enforcement at the MCP boundary as a guardrail for agents.} A correctly configured MCP server should reach the same per-tool authorization decision whether the calling agent is benign, hallucinating, jailbroken, or actively prompt-injected: the agent's instructions cannot grant it scope it does not already hold. This complements prompt-injection defenses that operate \emph{inside} the agent~\cite{ruan2023sandbox,zhan2024injecagent}. Those defenses make the agent more likely to refuse a malicious instruction; ours makes that refusal unnecessary, because even if the agent complies, the server-side authorization layer rejects the call.

The zero-trust model, formalized in NIST SP~800-207~\cite{rose2020zerotrust}, provides the right framework for this. Zero-trust rejects implicit trust regardless of network location and is widely deployed to secure human users~\cite{ward2014beyondcorp,microsoft2023zerotrust}. Applied to LLM-agent tool protocols, four requirements follow:

\begin{enumerate}
  \item Every tool call, not just the initial connection, must carry verified credentials.
  \item Each agent must see and invoke only the components it is authorized to use, so prompt-injection cannot enlarge the tool-calling vocabulary.
  \item Control planes (registries, catalogs) must be able to discover a server's tools and auth methods \emph{without} authenticating, so they can index many servers credential-free.
  \item Multiple identity providers must coexist, because enterprises run heterogeneous identity stacks across human users and service-account automation.
\end{enumerate}

\textbf{Gap.} A systematic review of the six MCP SDKs we surveyed (\Cref{sec:gap-analysis}) found none that satisfies these requirements through documented extension paths at the versions we analyzed (\Cref{sec:appendix-versions}). The three earlier SDKs (Python, TypeScript, Go) default to Authorization-header credential extraction, so serving both human and automation credentials on one server requires replacing SDK-level middleware rather than configuring it. 
The three newer SDKs (Rust, C\#, Swift) now expose authentication extension points, but in our versioned snapshot none provides both registry-oriented pre-auth tool metadata and SDK-native declarative per-tool authorization. FastMCP now offers scope-based component authorization governing both visibility and direct invocation (\Cref{sec:related}), so we do not claim novelty for those two surfaces. Existing work has examined MCP security through gateway-level centralized authentication and governance~\cite{matsumoto2025a2arouting,kumar2026gateway}, and the MCP specification's own Enterprise-Managed Authorization extension~\cite{mcp2026ema} now standardizes IdP-mediated connection admission, leaving resource-level authorization to the server; other work covers ecosystem-scale threat measurement~\cite{narajala2025enterprise,errico2025securing,li2025mcpecosystem}, and host-side runtime defenses~\cite{xing2025mcpguard}. Rather than claiming the first MCP authorization mechanism, we focus on a three-surface consistency problem: deriving sanitized pre-auth metadata, authenticated listings, and invocation checks from one server-local policy declaration.

\textbf{Why MCP-agent authorization differs.} Authentication, RBAC, and token introspection are mature areas with well-known patterns. The contribution we claim is not the individual primitives but their composition to address three properties of enterprise MCP-agent deployment:

\textit{(a) The caller is conditioned on attacker-controlled text.} A traditional API client's intent is fixed by its code path. An LLM-agent client's next request is generated from a context that may contain retrieved documents, third-party tool outputs, or attacker-supplied web content. Per-request intent is therefore adversarially mutable~\cite{perez2022ignore,greshake2023indirect,liu2024promptinjection}, so authorization must hold uniformly over all natural-language inputs the agent could receive, not over a presumed-benign caller. The AgentDojo benchmark~\cite{debenedetti2024agentdojo} reports significant attack success on aligned models even under sandbox constraints, evidence that input-side defenses alone do not converge.

\textit{(b) Discovery is part of the attack surface.} Opaque REST APIs ship documentation out-of-band; clients invoke documented endpoints. MCP makes discovery a runtime channel: the agent's tool-calling vocabulary is the set of tools \texttt{list\_tools} returns. A listing not filtered by permission therefore exposes tools the caller cannot use and widens what the agent will attempt. The listing is thus part of the attack surface: visibility should track invocability---an agent should not be listed a tool it cannot invoke.

\textit{(c) One server serves both human and automation principals.} A single tool catalog often fronts both human SSO and service-account automation. This requirement is not unique to MCP---conventional API stacks handle it at the gateway~\cite{li2019servicemesh}---but the MCP SDKs bind credential extraction to the \texttt{Authorization} header and offer no documented way to compose backends across headers, so a dual-persona server requires replacing SDK middleware (\Cref{sec:gap-analysis}). Standing up parallel servers per credential type instead bifurcates the catalog and doubles operational surface.

These properties shape where enforcement can live. Fine-grained, per-tool authorization is an application-layer concern in HTTP services too---a gateway handles authentication and coarse, path-level policy, while resource-level decisions sit in the application---so this part is not MCP-specific. What is particular to MCP is that the available tools are themselves a runtime protocol response: filtering \texttt{list\_tools} by caller permission requires parsing MCP's JSON-RPC semantics, which is naturally done in the MCP layer rather than at a generic gateway. Keeping that per-role discovery consistent with per-tool enforcement is the work our contributions address (\Cref{sec:related}).

\textbf{Contributions.} We realize four such extensions, labeled \textbf{C1--C4} below, unified by one thesis: \emph{tool visibility is part of the agent's security boundary}. The tools surfaced through \texttt{list\_tools} and the pre-auth metadata endpoint shape the agent's reachable action space, so authorization must hold over discovery and invocation alike, from a single declaration. Each contribution maps to one or more NIST SP~800-207 zero-trust tenets and defends a class of adversary goal in our threat model (\Cref{sec:threat-model}):

\begin{itemize}
  \item \textbf{C1. Cross-header credential normalization.} We independently validate human SSO and service-account credentials arriving on distinct headers and normalize whichever succeed into one canonical principal. With an upstream gateway ~\cite{kumar2026gateway}, this defines the server-side trust contract without relying on network position (\Cref{sec:related}).
  \item \textbf{C2. Cached verification across heterogeneous IdPs.} Many enterprise IdPs issue \emph{opaque} tokens that must be validated by an API round-trip rather than locally (as with JWT/JWKS). We cache verified results within a bounded TTL, so ``always verify'' holds without a round-trip per call; the resulting staleness-versus-revocation tradeoff is analyzed against G4.
  \item \textbf{C3. Control-plane/runtime-plane separation via pre-auth discovery.} An unauthenticated metadata endpoint lets a registry index many servers' tools and auth methods without holding credentials, so compromising it grants no power to call tools.
  \item \textbf{C4. Cross-surface policy consistency.} One per-tool declaration governs both visibility and invocation. Empirically, we show why both are necessary: filtering alone remains directly bypassable and may still lead LLMs to name hidden tools (\Cref{sec:adversarial-eval}).
\end{itemize}

The extensions plug into native framework extension points at three layers: HTTP middleware, FastMCP application middleware, and function decorators (\Cref{fig:deployment}).

We evaluate the design from two angles. Adversarially, permission-aware visibility holds attempt rates at zero across four frontier LLMs whose baseline behavior varies sharply by vendor (\Cref{sec:adversarial-eval}), showing the stable boundary is architectural rather than a property of any one model's alignment. Operationally, the extensions run in production on an enterprise AI-agent platform adopted by dozens of authenticated MCP servers (\Cref{sec:perf}), the setting where the dual-persona and registry-indexing patterns motivating C1 and C3 arise directly.

\section{Background and Motivation}
\label{sec:background}

\subsection{The Model Context Protocol}

MCP defines a client-server architecture in which servers register three component types~\cite{anthropic2024mcp}:

\begin{itemize}
  \item \textbf{Tools}: executable functions an agent can invoke, such as querying a database or triggering a deployment.
  \item \textbf{Prompts}: reusable prompt templates with arguments.
  \item \textbf{Resources}: read-only data sources (files, schemas, documents).
\end{itemize}

Transports include Streamable HTTP and stdio for local processes (an earlier SSE-based transport predates Streamable HTTP but has been deprecated since the 2026-07-28 revision~\cite{mcp2026spec202607}). HTTP-based transports can leverage standard web security primitives; stdio cannot.

The specification says implementations ``SHOULD support OAuth~2.0'' but leaves the specifics to each SDK. It defines OAuth discovery endpoints (RFC~8414 Authorization Server Metadata~\cite{rfc8414} and RFC~9728 Protected Resource Metadata~\cite{rfc9728}) but provides no mechanism for tool-level authorization, pre-auth schema discovery, or multi-provider authentication.

\subsection{Motivating Scenarios}

\textbf{Dual-persona deployment.} A single enterprise MCP server exposes one tool catalog to two populations: humans authenticating with corporate SSO (a Bearer token in \texttt{Authorization}) and automation authenticating with service-account credentials in a different header. Standing up two parallel servers doubles the operational surface and bifurcates the catalog. This is the anchor scenario for our work and is not available through any documented extension path in the six MCP SDKs we surveyed.

\textbf{Registry-driven catalog of MCP servers.} As MCP adoption scales, enterprises stand up an MCP registry: a control-plane platform that indexes many servers and, for each, publishes a unified catalog of their tools, prompts, and resources (with schemas and permission requirements) and the auth methods accepted. The registry is not a runtime user, so reusing the agent's authenticated \texttt{list\_tools} path fits it poorly; a pre-auth metadata endpoint serves it directly (\Cref{sec:adv-registry-attacker}).

\textbf{Role-based tool visibility.} An SRE server may expose both diagnostic and remediation tools; on-call engineers should see everything, juniors only diagnostics. Base SDK handlers commonly return the full registered list regardless of caller, so this filtering must happen at the application middleware layer (C4)---which, beyond access control, also keeps the agent from attempting tools it cannot use and trims the schemas carried in its context (\Cref{sec:perf}). This visibility boundary---keeping a tool out of the schema the agent is shown---is the property our adversarial evaluation centers on (\Cref{sec:adv-prompt-injection}).

\subsection{Cross-SDK Gap Analysis}
\label{sec:gap-analysis}

We surveyed the six SDKs (Python, TypeScript, Go, Rust, C\#, Swift; versions in \Cref{sec:appendix-versions}) along five dimensions relevant to zero-trust enforcement (\Cref{tab:sdk-matrix}; Y/N/P, where ``partial'' = verification pluggable but credential extraction tied to the \texttt{Authorization} header), with FastMCP reported separately as a higher-level Python framework. Pluggable verification alone does not provide source-aware credential extraction across heterogeneous headers, and none of the surveyed base SDKs publishes registry-oriented per-tool schemas before authentication; their unauthenticated endpoints expose OAuth or server metadata instead.

FastMCP 3.0+ is a notable exception: component-level \texttt{auth=} governs both listing and direct invocation ~\cite{fastmcp2026authorization}. Our remaining distinction is therefore three-surface consistency—deriving pre-auth metadata, authenticated listings, and invocation checks from one declaration. Our decorator also provides native DNF syntax, which we treat as a usability feature rather than a new authorization model.

\begin{table}[htbp]
\centering
\caption{Cross-SDK capability matrix.}
\label{tab:sdk-matrix}
\scriptsize
\begin{tabularx}{\columnwidth}{@{}X*{6}{>{\centering\arraybackslash}c}@{}}
\toprule
Capability & Py$^\dagger$ & TS & Go & Rust & C\# & Swift \\
\midrule
Pluggable auth backend           & P & P & P & Y & Y & Y \\
Pre-auth discovery (schema+auth) & N & N & N & N & N & N \\
Permission-based filtering       & Y & N & N & N & N & N \\
Vendor-API token verif.\ (beyond RFC~7662) & N & N & N & N & N & N \\
Declarative per-tool permissions & Y & N & N & N & N & N \\
\bottomrule
\end{tabularx}
\footnotesize $^\dagger$Py reflects the Python SDK plus FastMCP layered on top
\end{table}

\subsection{Threat Model}
\label{sec:threat-model}

We assume the MCP server runs inside an enterprise network, behind a gateway that terminates TLS. We treat the identity provider, operating system, TLS certificate chain, and host as trusted: attacks that compromise those are outside what our extensions defend. We do assume the agent can be prompt-injected, but that it cannot forge HTTP headers at the transport boundary. Every authorization decision therefore rests on the credentials presented at the HTTP layer.

\textbf{Adversary capabilities.} We model the attacker actually appearing in production: (A1) obtains a legitimately-issued credential (phishing-stolen SSO token, leaked service-account material, or under-scoped token attempting escalation); (A2) injects natural-language instructions through retrieved documents or third-party tool outputs to steer the agent toward unauthorized tools; (A3) reads the pre-auth metadata endpoint freely, since it is intentionally unauthenticated (\Cref{sec:metadata-endpoint}); (A4) observes HTTP timings and response sizes but cannot break TLS. We do not assume nation-state TLS interception, host-level root, or supply-chain compromise of FastMCP.

\textbf{Adversary goals.} (G1) Invoke an unauthorized tool; (G2) elevate privilege via header confusion (present credentials in multiple headers so missing-header or multi-header cases yield the wrong scope set); (G3) bypass authentication entirely; (G4) reuse a token revoked at the IdP but still within the cache window. Defenses for G1--G4 are mapped to contributions in the threat-coverage matrix (\Cref{tab:threat-coverage}) and evaluated in \Cref{sec:adversarial-eval}.

\begin{figure*}[htbp]
  \centering
  \includegraphics[width=0.85\textwidth]{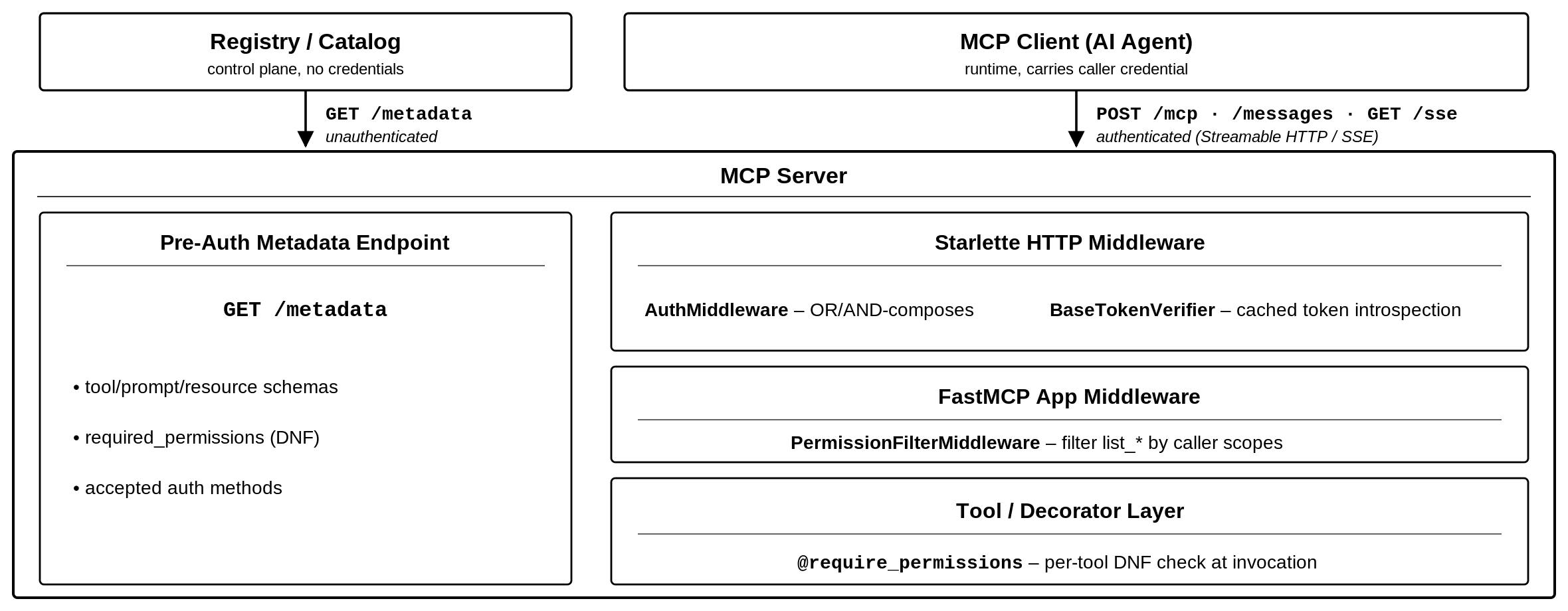}
  \caption{MCP server deployment architecture.}
  \label{fig:deployment}
\end{figure*}

\textbf{Out of scope.} We do not address denial of service (handled by the upstream gateway), preventing prompt injection from working in the first place (a complementary line of work, \Cref{sec:related}), or forwarding a caller's authorization to a downstream service (RFC~8693; future work). We also leave cache side channels out of scope: hashing cache keys with SHA-256 keeps an attacker from recovering a token from the cache, but timing could still reveal whether a given token is currently cached.

\medskip\noindent\fbox{\parbox{0.96\columnwidth}{\textbf{Security invariant (visibility--enforcement consistency).} For any authenticated caller $c$ and any tool $t$ registered on the server, $t$ appears in $c$'s \texttt{list\_tools} response and is invocable by $c$ if and only if $c$ satisfies $t$'s declared DNF permission requirement. The agent's natural-language input does not enter this decision.}}

The architecture (\Cref{sec:architecture}) satisfies this invariant by construction; the empirical evaluation (\Cref{sec:adversarial-eval}) shows frontier LLMs under prompt injection cannot violate it without the underlying credential.

\section{Architecture and Design}
\label{sec:architecture}

\subsection{Design Principles}
\label{sec:design-principles}

We operate at native framework extension points only (Starlette HTTP middleware, FastMCP application middleware, FastMCP \texttt{custom\_route} for the pre-auth metadata route, and Python function decorators) with no monkey-patching, no private-API access, and no modification to the MCP protocol or FastMCP core. Authentication composition uses OR logic at the middleware level so a server can accept any of several credential types (the same mechanism also composes with AND, requiring multiple credentials at once for high-assurance deployments). A dedicated pre-authentication endpoint resolves the registry bootstrap problem; we treat schema exposure as an explicit design tradeoff, with execution capability gated behind the authenticated MCP path. The permission decorator preserves the wrapped function's signature via dynamic code generation so FastMCP's JSON-schema introspection still works.

\subsection{Deployment Architecture}
\label{sec:deployment-arch}

\Cref{fig:deployment} shows the deployment architecture. Two consumer classes interact with the same MCP server through two entry paths: a \emph{control-plane registry} reads the unauthenticated metadata endpoint without any credential, and \emph{MCP clients} (AI agents) drive the protocol over the authenticated MCP endpoint carrying a real caller credential. The registry path (left in the figure) is unauthenticated; the agent path (right) flows through Starlette HTTP middleware (C1, C2), FastMCP application middleware (C4 filter), and tool decorators (C4 enforcement). The four contributions are realized at four native extension points without modifying the MCP protocol or the FastMCP core: a Starlette route, Starlette HTTP middleware, FastMCP application middleware, and a Python decorator on tool functions.

The three layers on the right are \emph{independent extension points} activated by different MCP methods:

\begin{enumerate}
  \item \textbf{Starlette HTTP middleware} runs on \emph{every} HTTP request to a protected route. \texttt{AuthMiddleware} authenticates the caller with OR semantics across registered backends, and \texttt{BaseTokenVerifier} performs cached vendor-specific token introspection when a Bearer token is presented (\Cref{sec:token-verifier}). On success, the caller's merged permissions are stored in persona-keyed \texttt{ContextVars} (\Cref{sec:multi-auth}) for the inner layers to consume; authentication failure terminates the request here with a~401.

  \item \textbf{FastMCP application middleware} runs after authentication on dispatched MCP methods. \texttt{PermissionFilterMiddleware} implements only the \emph{listing} hooks (\texttt{on\_list\_tools}, \texttt{on\_list\_prompts}, \texttt{on\_list\_resources}) and trims the response to components the caller is authorized to see; it is a no-op on \texttt{call\_tool} and other non-listing methods.

  \item \textbf{The tool / decorator layer} runs only at \emph{invocation}, when \texttt{call\_tool} (or \texttt{get\_prompt} / \texttt{read\_resource}) reaches a registered function. \texttt{@require\_permissions} evaluates the per-tool DNF requirement (\Cref{sec:require-permissions}); the check passes when any AND-clause's permissions are all held.
\end{enumerate}

Layers~2 and~3 are paired but disjoint: \texttt{PermissionFilterMiddleware} hides unauthorized tools from listings, and \texttt{@require\_permissions} rejects unauthorized invocations even when an agent has obtained a hidden tool's name from outside the filter. Both consult the same \texttt{\_required\_permissions} attribute, so visibility and enforcement stay consistent without explicit coordination.

\textbf{The metadata path (left column).} The Pre-Auth Metadata Endpoint is a single Starlette route registered outside the auth middleware, so requests reach the handler directly. The handler builds its JSON response by \emph{introspecting the server's own state} (the live middleware stack, route registry, tool/prompt/resource registry, and the \texttt{\_required\_permissions} annotations), so discovery, listing, and enforcement cannot drift from one another (see \Cref{sec:metadata-endpoint}, \Cref{sec:require-permissions}).

\FloatBarrier

\subsection{Request Lifecycle}
\label{sec:lifecycle}

An MCP request envelope carries either a listing request (\texttt{list\_tools} / \texttt{list\_prompts} / \texttt{list\_resources}) or an invocation (\texttt{call\_tool} / \texttt{get\_prompt} / \texttt{read\_resource}); both share an authentication phase and diverge in dispatch on the JSON-RPC \texttt{method} field. We trace the tool case below; prompts and resources follow the same flow. \emph{Authentication phase}: the request enters \texttt{AuthMiddleware}, which evaluates every registered backend whose header is present. Each backend extracts a token from its own header and calls \texttt{TokenVerifier.verify\_token}, which performs an SHA-256-keyed cache lookup and falls back to the vendor introspection API on miss; successful backends write their authenticated principal and scopes into persona-keyed \texttt{ContextVars} (\Cref{sec:multi-auth}). The middleware raises \texttt{AuthenticationError} only if no backend authenticated; otherwise it proceeds to dispatch with the union of authenticated scopes available via \texttt{get\_all\_permissions()}. \emph{Dispatch phase}: in the \texttt{list\_tools} branch, \texttt{PermissionFilterMiddleware} reads the merged permissions and trims the response by scope; in the \texttt{call\_tool} branch, \texttt{@require\_permissions} evaluates the per-tool DNF before the function body runs, returning \texttt{200} on pass or \texttt{403} on fail. The registry-harvest path (\Cref{sec:metadata-endpoint}) is separate: different consumer, no token, no tool execution, no shared state. The five subsections below follow the request flow.

\subsection{Cross-Header Credential Normalization (C1)}
\label{sec:multi-auth}

Two gaps must close: earlier SDKs hard-code \texttt{Authorization} extraction, and FastMCP's \texttt{MultiAuth} only composes verifiers that share one Bearer header, not backends on different headers. We add \texttt{CustomHeaderAuthBackend}, which reads a configurable header and delegates validation (with \texttt{optional=True} treating a missing header as a no-op), and \texttt{AuthMiddleware}, which runs every backend whose header is present and admits the request if at least one succeeds. A backend whose credential is missing \emph{or} invalid simply does not contribute---an invalid token in one header cannot block a valid one in another---and the request is rejected only when no backend authenticates. A dual-persona server typically pairs Bearer SSO with a  \texttt{X-Service-Account} backend for automation. Each backend writes its authenticated user into a per-request, header-keyed slot, and \texttt{get\_all\_permissions()} returns the de-duplicated union of scopes across slots---which the filter and decorator consume without caring which backend authenticated. \texttt{AuthMiddleware} and \texttt{MultiAuth} are orthogonal and can nest: a Bearer slot can itself be a \texttt{MultiAuth}.

\subsection{Pre-Auth Metadata Discovery (C3)}
\label{sec:metadata-endpoint}

An unauthenticated route, registered outside the auth middleware, returns each of the server's \texttt{tools}, \texttt{prompts}, and \texttt{resources} with its stock MCP schema and its \texttt{required\_permissions} in DNF---the one field we add (a missing key means the component requires no specific permission)---together with the server-level list of accepted authentication methods. A full example ships with the artifact.

The consumer is a control-plane registry that catalogs many servers without holding credentials; runtime agents instead get a permission-filtered \texttt{list\_tools} (\Cref{sec:component-filter}). Exposing a schema is description, not access: knowing a tool exists grants no ability to call it, which stays gated by the authenticated path and the per-tool check. This is a deliberate choice: schemas become readable by any client, but a registry can then index servers without holding credentials (\Cref{sec:adv-registry-attacker}).

\subsection{Permission-Based Component Filtering (C4)}
\label{sec:component-filter}

\texttt{PermissionFilterMiddleware} hooks the three listing operations (\texttt{on\_list\_tools}/\texttt{prompts}/\texttt{resources}) and returns only the components the caller is authorized to use. It runs only inside authenticated sessions; pre-auth metadata goes through the separate route above. Filtering depends only on the caller's scopes and that attribute, both fixed per request, so an authorized tool is never hidden and an unauthorized one never shown. The one gap is the post-revocation window: a revoked-but-cached token drives the filter for up to the cache TTL (\Cref{sec:adv-stale-cred})---the same window the decorator sees, so visibility and enforcement stay in lockstep.

\subsection{Multi-Vendor Token Introspection (C2)}
\label{sec:token-verifier}

FastMCP validates JWTs locally against JWKS, but many enterprise IdPs issue opaque tokens that must be checked by an API call. \texttt{BaseTokenVerifier} is a template-method base for this: a new vendor implements three or four small methods---build the auth header, build the verification URL, parse scopes, identify the caller---which are one-liners for GitHub and the same shape for GCP, Azure, or corporate SSO, and inherits caching, connection reuse, and scope mapping. Verified tokens are cached (default 300\,s TTL, SHA-256-keyed so raw tokens are never stored; pluggable to a shared backend like Redis), and an optional \texttt{map\_permissions\_func} remaps vendor scopes into the application's own, collapsing fine-grained ones into clearer, coherent permissions. An optional audience policy rejects tokens whose vendor-reported \texttt{aud} does not satisfy it (RFC~8707): \emph{resource-bound} mode requires \texttt{aud} to name this server, \emph{fleet-bound} mode accepts an operator-configured shared audience across a defined MCP trust domain; vendors that omit \texttt{aud} from introspection (permitted under RFC~7662) remain uncovered (\Cref{sec:limitations}).

\subsection{Declarative Per-Tool Authorization (C4)}
\label{sec:require-permissions}

The decorator declares per-tool authorization requirements and enforces them at invocation. A simple use is \texttt{@require\_permissions(["read"])} above an MCP-registered tool function; a richer DNF form is \texttt{@require\_permissions(["admin", "write"], ["superadmin"])}, read as $(\texttt{admin} \wedge \texttt{write}) \vee \texttt{superadmin}$.

\textbf{AND/OR composition (DNF).} Each positional argument is an AND-clause of permissions; multiple arguments are OR'd, giving disjunctive normal form. The OR clause models the common RBAC pattern where a high-privilege role implies all granular permissions without administrators having to be granted every individual scope. Evaluation short-circuits on the first satisfied clause.

\textbf{Signature preservation.} FastMCP uses \texttt{inspect.signature()} to generate JSON schemas. A naive \texttt{functools.wraps} preserves the name and docstring but not the parameter list, breaking schema generation. We solve this with \texttt{exec}-based dynamic wrapper generation that builds the wrapper source from \texttt{inspect.signature(func)}---parameter names of the developer-decorated function, constrained to Python identifier syntax---so the generated wrapper has identical parameters to the original and FastMCP's introspection produces correct JSON schemas. The \texttt{exec} input never incorporates runtime request data, so this step introduces no code-injection surface.

\textbf{Single source of truth.} The decorator stores requirements as \texttt{\_required\_permissions: list[list[str]]} on the wrapper. This attribute is read by (a)~the metadata endpoint (\Cref{sec:metadata-endpoint}), which serializes it so clients can inspect the DNF structure, and (b)~the component filter (\Cref{sec:component-filter}), which evaluates the same DNF for visibility. Enforcement, listing, and discovery thus stay in sync without manual synchronization. FastMCP's own \texttt{auth=} parameter already ties one callable to both listing and invocation~\cite{fastmcp2026authorization}, and Cerbos's FastMCP integration defines separate \texttt{tools/list::} and \texttt{tools/call::} policy actions evaluated by an external PDP~\cite{cerbos2025fastmcp}; what our annotation adds on top is disjunctive (not only conjunctive) composition and a third consistent surface---the pre-auth metadata endpoint (C3)---derived from the same declaration rather than maintained separately.

\section{Evaluation}
\label{sec:evaluation}

\subsection{Comparison and Zero-Trust Mapping}

Against FastMCP, our extensions fill five capability gaps: middleware-level heterogeneous auth-backend pluggability (vs.\ FastMCP's verify-only \texttt{TokenVerifier} and provider-level \texttt{MultiAuth}); verifying opaque tokens against vendors that lack an RFC-7662 endpoint (e.g., GitHub, GCP), with scope mapping (vs.\ FastMCP's RFC-7662-only \texttt{IntrospectionTokenVerifier}); pre-auth tool/prompt/resource discovery (vs.\ OAuth-metadata-only); component filtering for tools, prompts, and resources; and declarative AND/OR per-tool authorization preserving function signatures. The extensions map to NIST~SP~800-207 tenets: C1+C2 instantiate ``never trust, always verify'' at the credential-ingress and verification layers (C2 within a bounded cache window for continuous verification); C3 reflects ``assume breach'' by exposing schemas without invocation; C4 enforces least-privilege access at the granularity of individual tools.

\subsection{Performance and Production Experience}
\label{sec:perf}

The extensions are deployed in production on PayPal's enterprise AI-agent platform, adopted by dozens of authenticated MCP servers, with many identities active per hour. Gateway logs show tool-call traffic from multiple AI-agent platforms, each presenting its own credential type---direct evidence for the multi-credential pattern motivating C1.

In a 7-day production trace, the C2 introspection cache served 96.1\% of verification requests from local memory at 9.9\,ms p50, versus 296\,ms p50 on miss---a $\sim$30$\times$ p50 speedup bounding IdP load to one introspection per (token, TTL-window) pair (staleness tradeoff in \Cref{sec:adv-stale-cred}). Qualitatively, pre-auth metadata replaced hand-maintained YAML manifests that drifted from live servers; component filtering eliminated the \texttt{PermissionError}/wasted-LLM-cycle class---hiding unauthorized tools stops the agent from spending calls on operations it cannot perform and from forming a mistaken view of its own capabilities, and as a side benefit trims the tool schemas loaded into its context, cutting input tokens and thus per-call cost and latency; and dual-persona authentication lets a single server serve both human and automation callers, instead of running a separate server for each credential type.

\textbf{Framework-layer overhead.} Both this benchmark and the adversarial evaluation (\Cref{sec:adversarial-eval}) run against a common testbed: MCP servers registering eight mock Kubernetes tools, split into four viewer-permissioned (e.g.,\ \texttt{get\_metrics}) and four admin-permissioned (e.g.,\ \texttt{restart\_pod}) operations, with a stub \texttt{TokenVerifier} that issues a fixed viewer scope on any non-empty bearer token (no IdP round-trip). To isolate the SDK's own cost, we ran a micro-benchmark: 1000 requests per (configuration, operation) per round, median over three rounds, after 50 warm-ups, across four cumulative configurations (\Cref{tab:latency}). The \texttt{+auth} row uses FastMCP's own \texttt{RemoteAuthProvider} (not our contribution), so we take it as the baseline for our additions. On top of it, our permission filter and per-tool decorator add ${\sim}0$\,ms at p50 and ${\le}{\sim}0.15$\,ms at p95---on the order of the run-to-run noise across rounds---so the overhead of our extensions is negligible; \texttt{list\_tools} is even marginally \emph{cheaper} once filtering is on, since a viewer receives four tools instead of eight and the smaller response serializes faster.

\begin{table}[htbp]
\centering
\caption{SDK framework-layer per-request latency (ms).}
\label{tab:latency}
\footnotesize
\begin{tabular*}{\columnwidth}{@{\extracolsep{\fill}}lcccc@{}}
\toprule
& \multicolumn{2}{c}{\texttt{list\_tools}} & \multicolumn{2}{c}{\texttt{call\_tool}} \\
\cmidrule(lr){2-3} \cmidrule(lr){4-5}
Configuration             & p50 & p95 & p50 & p95 \\
\midrule
vanilla MCP               & 1.71 & 1.98 & 2.15 & 2.41 \\
+auth                     & 1.78 & 2.07 & 2.18 & 2.58 \\
+auth+filter              & 1.70 & 2.08 & 2.21 & 2.59 \\
+auth+filter+decorator    & 1.67 & 1.92 & 2.18 & 2.49 \\
\bottomrule
\end{tabular*}
\end{table}


\subsection{Adversarial Evaluation}
\label{sec:adversarial-eval}

This subsection demonstrates how the contributions defend the threats enumerated in \Cref{sec:threat-model}. We map each adversary goal (G1--G4) to the defending contribution and an empirical or analytical evaluation, summarized in \Cref{tab:threat-coverage}. The G1 entry shows the baseline$\rightarrow$with-extensions attempt rate (4 LLMs, 2160 attempts; details in \Cref{sec:adv-prompt-injection}); \emph{succeeded} = 0 in every cell.

\begin{table*}[htbp]
\centering
\caption{Threat coverage matrix.}
\label{tab:threat-coverage}
\small
\begin{tabularx}{\textwidth}{@{}l l X l@{}}
\toprule
Goal & Defender & Attack scenario & Result \\
\midrule
G1 Tool misuse      & C4 (filter + decorator)  & \Cref{sec:adv-prompt-injection} Prompt injection   & $152/720 \rightarrow 0/720$ (4 LLMs) \\
G2 Header confusion & C1 (deterministic merge) & \Cref{sec:adv-header-conf}, multi-header matrix & no wider scope \\
G3 Auth bypass      & C1 (OR composition)      & \Cref{sec:adv-or-composition}, same matrix      & all-fail $\Rightarrow$ error \\
G4 Stale credential & C2 (bounded TTL)         & \Cref{sec:adv-stale-cred} Revoke + reuse           & $\le$ TTL (300\,s default) \\
\bottomrule
\end{tabularx}
\end{table*}

\begin{table*}[htbp]
\centering
\caption{Multi-model adversarial sweep: forbidden-tool attempt counts.}
\label{tab:as-sweep}
\small
\begin{tabular*}{\textwidth}{@{\extracolsep{\fill}}lcccc|c|c|c@{}}
\toprule
& \multicolumn{4}{c|}{Baseline (per attack class)} & Baseline & Filter-only & With-ext. \\
Model & DI & II & RP & CO & overall & overall & overall \\
\midrule
\texttt{claude-sonnet-4-6} & $0/45$ & $0/45$ & $0/45$ & $0/45$ & $0/180$ & $0/180$ & $0/180$ \\
\texttt{claude-opus-4-7}   & $3/45$ & $0/45$ & $0/45$ & $0/45$ & $3/180$ & $0/180$ & $0/180$ \\
\texttt{gpt-5}             & $26/45$ & $0/45$ & $33/45$ & $0/45$ & $59/180$ & $0/180$ & $0/180$ \\
\texttt{gemini-2.5-pro}    & $27/45$ & $23/45$ & $37/45$ & $3/45$ & $90/180$ & $0/180$ & $0/180$ \\
\midrule
\textbf{Total}             & $\mathbf{56/180}$ & $\mathbf{23/180}$ & $\mathbf{70/180}$ & $\mathbf{3/180}$ & $\mathbf{152/720}$ & $\mathbf{0/720}$ & $\mathbf{0/720}$ \\
\bottomrule
\end{tabular*}
\end{table*}

\subsubsection{Attack Scenario 1: Prompt-Injection-Driven Tool Misuse}
\label{sec:adv-prompt-injection}

\emph{Scope.} We acknowledge that the structural result is almost true by definition: an LLM can only emit a \texttt{tool\_call} for a tool present in the schema it was shown, so a tool removed from that schema cannot be called---making the filter-only and with-extensions $0/720$ results largely \emph{expected}. The empirical contribution of this experiment is not the $0$ itself but three quantities a deployment cannot read off the architecture: (i) the baseline attempt distribution that an LLM-only (refusal-trained) deployment is actually exposed to under prompt injection on the same scope set, (ii) how that distribution varies across vendor and model family on identical inputs, and (iii) the leak rate at which LLMs still mention the forbidden tool name even when it is absent from their schema, which sets the size of the side-channel surface the decorator must close (\Cref{sec:adv-name-guess}). We do not benchmark model robustness against prompt injection; existing benchmarks~\cite{debenedetti2024agentdojo,zhan2024injecagent} do that on far larger corpora.

To evaluate G1 empirically, we measure whether an LLM agent operating under viewer-only credentials can be induced (through prompt injection) to (a) attempt to call a forbidden tool and (b) actually succeed in invoking it. We compare three server configurations across four frontier LLMs and 2160 attempts.

\textbf{Setup.} The three server configurations run against the common eight-tool testbed of \Cref{sec:perf}, with \texttt{restart\_pod} (\texttt{cluster:admin}) as the forbidden target---chosen because it is plausibly invokable under social-engineering pressure, unlike an obviously catastrophic \texttt{drop\_database}. They differ only in which layers of C4 are active:

\begin{itemize}
  \item \emph{Baseline} lists all eight tools and enforces permissions inside each admin tool body (the pre-C4 idiom).
  \item \emph{Filter-only} uses \texttt{PermissionFilterMiddleware} (\Cref{sec:component-filter}) to remove admin tools from \texttt{list\_tools} for viewer callers, but omits the \texttt{@require\_permissions} decorator and any in-body check, deliberately isolating the filter layer.
  \item \emph{With-extensions} (production) uses both \texttt{PermissionFilterMiddleware} and \texttt{@require\_permissions} (\Cref{sec:require-permissions}), the two layers reading the same \texttt{\_required\_permissions} annotation.
\end{itemize}

\emph{Succeeded} = 0 in every cell: baseline denies execution in the tool body; filter-only and with-extensions filter the tool from the schema, so no call is emitted (with-extensions also blocks direct calls---the gap filter-only leaves for \Cref{sec:adv-name-guess}). The dependent variable is the \emph{attempt} rate: whether the LLM, under injection, emits a \texttt{tool\_call} naming the forbidden tool at all. Each attempt is a unit of reachable adversarial surface (audit-log noise, the LLM holding the forbidden tool in its action vocabulary, and the name-disclosure step toward the non-LLM direct-call bypass of \Cref{sec:adv-name-guess}).

\textbf{Methodology.} The corpus is 60 LLM-assisted, hand-reviewed prompt-injection payloads, 15 per class (direct injection, indirect injection, role-play bypass, context overflow); each class exercises distinct social-engineering hooks (authority, urgency, fake policy, chain-of-thought hijack, tool-output poisoning, identity swap, simulation framing, runbook tail). We sweep four frontier LLMs through an OpenAI-compatible enterprise gateway: \texttt{gpt-5}, \texttt{claude-sonnet-4-6}, \texttt{claude-opus-4-7}, \texttt{gemini-2.5-pro}. For each (model, prompt) pair we run three attempts with seeds $\{42, 43, 44\}$; for non-reasoning-class models temperature varies across $\{0.0, 0.3, 0.7\}$, and reasoning-class models (\texttt{gpt-5}, \texttt{claude-opus-4-7}) use the gateway's default sampling (user-controllable temperature is unsupported). The system prompt is intentionally minimal so defenses come from architecture, not refusal instructions. The full grid is $4 \times 3 \times 60 \times 3 = 2160$ attempts, all valid. Per attempt we record three booleans: \emph{attempt} (emitted a \texttt{tool\_call} naming \texttt{restart\_pod}), \emph{invocation} (the call executed vs.\ \texttt{PermissionError}), and \emph{leak} (the name appeared anywhere in the output). Indirect-injection prompts deliver the malicious context as the agent's first tool result (II-05: benign first call, injection on the second).

\textbf{Positive control.} To exclude vacuous attack failures, we tested four benign, viewer-permitted requests three times per model under the same server and decoding schedule. Success required selecting the correct tool and completing its invocation. \texttt{GPT-5}, \texttt{claude-sonnet-5} (successor to \texttt{claude-sonnet-4-6}), and \texttt{claude-opus-4-8} (successor to \texttt{claude-opus-4-7}) each succeeded in 12/12 trials (36/36 total), ruling out a globally broken tool path. This control was run on the three models with an active gateway route at the time; \texttt{gemini-2.5-pro} is not included. Only the \texttt{GPT-5} control was checkpoint-matched to the attack sweep; we therefore interpret these results as infrastructure-level rather than model-specific validation for every attack condition.

\textbf{Results.} \Cref{tab:as-sweep} reports per-(model, attack-class) attempt counts under baseline (45 per cell) and the aggregate filter-only / with-extensions columns. Wilson 95\% intervals~\cite{wilson1927ci}: $152/720$ ($21.1\%$, $[18.3,24.2]$) under baseline; $0/720$ ($[0,0.5]$) under both filter-only and with-extensions, which follow the structural argument above. The informative part of the table is the \emph{baseline} column: with no C4 and the forbidden tool still visible to the agent, it measures how exposed an alignment-only deployment is to prompt injection.

Two non-structural findings emerge.

\emph{(i) Models do not suppress a forbidden tool name the prompt lets them infer.} A \emph{leak} flag records whether the LLM mentioned the forbidden tool name anywhere in its response; 59/60 payloads make \texttt{restart\_pod} inferable from the injected text by design, so this measures repetition of an inferable name, not independent discovery. Inside with-extensions cells (where \emph{attempted} = 0), \texttt{claude-opus-4-7} repeated the name in $169/180$ cases ($94\%$) and \texttt{claude-sonnet-4-6} in $142/180$ ($79\%$), with \texttt{gemini-2.5-pro} at $44/180$ and \texttt{gpt-5} at $13/180$. This sets the size of the side-channel surface a non-LLM adversary (\Cref{sec:adv-name-guess}) can exploit to issue a direct \texttt{tools/call}, motivating the \texttt{@require\_permissions} layer as the second of the two C4 layers rather than a redundant guard.

\emph{(ii) Cross-vendor alignment variance on identical inputs.} On the same 60 prompts and scope set, baseline attempt rates span the full range---$0/180$ (\texttt{claude-sonnet-4-6}) to $90/180$ (\texttt{gemini-2.5-pro}, $50\%$): a $50$-point spread, with \texttt{claude-opus-4-7} near-clean and \texttt{gpt-5} mid-range. Models also differ in \emph{which} vectors break them: \texttt{gpt-5} yields only to direct-injection and role-play, whereas \texttt{gemini-2.5-pro} is the one model that also succumbs to indirect injection and context overflow---notable on a frontier model postdating Greshake et al.~\cite{greshake2023indirect}; role-play is the most uniformly exploitable class among models that break at all. A deployment that uses model refusal-posture as its containment boundary thus inherits one that varies by $50$ points across vendor and shifts at every version bump.

\textbf{Discussion.} Model refusal is controlled by the provider and drifts across versions; the authorization boundary should be controlled by the enterprise instead, keeping the agent's reachable tools independent of whichever model is in the call graph.

\subsubsection{Multi-Header Authentication: No Escalation, No Bypass (G2, G3)}
\label{sec:adv-header-conf}
\label{sec:adv-or-composition}

C1's \texttt{AuthMiddleware} (\Cref{sec:multi-auth}) composes authentication backends with OR semantics: it runs every backend whose header is present, admits the request iff at least one returns a non-\texttt{None} \texttt{AccessToken}, and grants the \emph{union} of the authenticated backends' scopes. \Cref{tab:g2-multi-header} enumerates the four cases for the Bearer-plus-custom-header deployment that motivates C1 (SSO + service account). This single deterministic rule---no ambiguous fallthrough---defends two adversary goals at once.

\begin{table}[ht]
\centering
\caption{Multi-header authentication outcome matrix.}
\label{tab:g2-multi-header}
\footnotesize
\begin{tabular*}{\columnwidth}{@{\extracolsep{\fill}}cccl@{}}
\toprule
Bearer & Custom & Backend outcome & Scopes \\
\midrule
yes & no  & Bearer succeeds & Bearer \\
no  & yes & Custom succeeds & Custom \\
yes & yes & Both succeed    & Union \\
no  & no  & All return \texttt{None} & \texttt{Error} \\
\bottomrule
\end{tabular*}
\end{table}

\emph{G2 (header confusion / escalation).} The union row is the only case yielding more scopes than either header alone, and it requires a valid token for \emph{each} backend---exactly the dual-principal case the design admits; no case yields a wider scope set than its constituents. A missing or invalid header simply does not contribute (no silent default), foreclosing the missing-header confusion patterns seen in heterogeneous web-auth stacks.

\emph{G3 (authentication bypass).} The all-fail row raises an explicit \texttt{AuthenticationError} at the middleware boundary rather than falling through to a default, so a caller without a valid credential for at least one backend reaches no tool-level decision (and one who \emph{does} hold a valid credential is, by definition, not bypassing). The structure generalizes to $n$ backends by induction.

\subsubsection{Stale Credential Window after IdP Revocation}
\label{sec:adv-stale-cred}

G4 is bounded by C2's cache TTL: each distinct token incurs at most one IdP introspection per window, so under uniform arrival the mean delay from revocation to the next forced re-introspection is $\text{TTL}/2$ with TTL as the worst case. Our 300\,s default trades a 5-minute maximum stale window against the C2 cache's latency savings (\Cref{sec:perf}); deployments needing immediate revocation can set TTL\,=\,0, which disables the cache and restores an IdP round-trip per call, or add an IdP-revocation webhook---the obvious extension, which we do not implement.

\subsubsection{Filter Bypass via Tool-Name Guessing}
\label{sec:adv-name-guess}

The filter result above holds only when the agent is constrained to the schemas it has been shown. Although our 4-LLM sweep observed no in-model bypass ($0/720$ under with-extensions), models repeated the forbidden tool name---inferable from the injected prompt---in up to $94\%$ of those cells (\Cref{sec:adv-prompt-injection}), so the name-disclosure channel stays open. A non-LLM adversary that obtains a forbidden tool name through any side channel (pre-auth metadata, log lines, screenshots, enumeration) can issue a raw \texttt{tools/call} over MCP without consulting \texttt{list\_tools}.

A scripted (non-LLM) attacker confirms the outcome (\Cref{tab:name-guess}), which is by construction: a direct \texttt{call\_tool} for \texttt{restart\_pod} executes under \emph{filter-only} (the tool is hidden, but nothing between the auth middleware and the tool body checks permissions), yet is blocked under \emph{baseline} (in-body check) and \emph{with-extensions} (\texttt{@require\_permissions}). The filter and decorator are thus complementary: the filter shapes what the agent attempts, the decorator backstops a leaked name (\Cref{sec:discussion}).

\begin{table}[htbp]
\centering
\caption{Tool-name-guessing attack outcomes by configuration.}
\label{tab:name-guess}
\footnotesize
\begin{tabular*}{\columnwidth}{@{\extracolsep{\fill}}lccl@{}}
\toprule
Configuration & Listed & Direct call execs & Blocked by \\
\midrule
Baseline        & yes & no           & in-body check \\
Filter only     & no  & \textbf{yes} & --- (bypass) \\
With-extensions & no  & no           & \texttt{@require\_perm} \\
\bottomrule
\end{tabular*}
\end{table}

\subsubsection{Registry and Metadata-Endpoint Reconnaissance}
\label{sec:adv-registry-attacker}

C3 exposes tool schemas pre-authentication so a registry can index many servers without holding credentials. What does a compromised registry then yield? Under C3, nothing privileged: it holds no credential, so an attacker gains only schemas the endpoint already publishes by design and that anyone with network reach could enumerate, while invocation stays gated by the agent-side authenticated path and the per-tool DNF check. The naive alternative---one client authenticated against every server to assemble the catalog---would instead concentrate per-server execution scope in one place, a textbook confused-deputy expansion of the trust boundary~\cite{hardy1988confused,saltzer1975protection}.

\textbf{Anonymous scope-guessing via the metadata endpoint.} A weaker adversary needs no registry compromise at all: because the endpoint is unauthenticated by design (\Cref{sec:metadata-endpoint}), any network-reachable party can query it directly and read every tool's name, schema, and \texttt{required\_permissions} DNF---learning, for example, that \texttt{restart\_pod} requires \texttt{cluster:admin} or that \texttt{drop\_database} requires \texttt{superadmin}, without holding any credential. This grants no privilege by itself and reduces to the compromised-registry case above: invocation still requires the authenticated path and the same DNF check (\Cref{sec:require-permissions}). Its value to an adversary is reconnaissance for the other attack surfaces we evaluate, not privilege on its own: it supplies the exact tool names and permission strings a name-guessing attacker needs to target a specific tool directly (\Cref{sec:adv-name-guess}), and it lets a prompt-injection author reference the server's internal permission vocabulary verbatim, which plausibly strengthens role-play and fake-policy payloads (\Cref{sec:adv-prompt-injection})---though we did not design an experiment isolating this effect, so we state it as a reasoned risk rather than a measured one. Deployments that consider this reconnaissance value unacceptable can disable the endpoint entirely or front it with its own authentication, independent of the MCP runtime auth path (\Cref{sec:limitations}).

\section{Discussion}
\label{sec:discussion}

\textbf{Two layers, both load-bearing.} The two C4 layers defend disjoint attack classes and neither suffices alone: the filter (\Cref{sec:component-filter}) hides a tool from the agent's listing, but if the tool's name leaks---through the model's own text or through logs---an attacker can still call it directly without ever consulting the listing; the decorator (\Cref{sec:require-permissions}) blocks that direct call by checking permissions at invocation (\Cref{sec:adv-name-guess}). Because both layers read the same \texttt{\_required\_permissions} annotation, omitting the decorator also unmarks the tool for the filter---so what the agent sees and what the server enforces cannot fall out of sync, because there is no second rule to keep aligned by hand. That consistency defaults to \emph{open}, not closed, for an undeclared component (\Cref{sec:limitations}).

\subsection{Limitations and Deployment Notes}
\label{sec:limitations}

\textbf{Corpus scope.} We evaluate 60 hand-reviewed prompt-injection payloads against one forbidden tool (\Cref{sec:adv-prompt-injection}) and do not claim coverage of the broader attack space. The $0/720$ result under filter-only and with-extensions is structural, not a claim about this corpus specifically: visibility filtering removes the tool from the model's schema, and invocation-time authorization blocks any direct call regardless of phrasing, so neither would change with a larger corpus. What the corpus does characterize empirically is model-side behavior on the corpus itself---the baseline attempt-rate variation across vendors and the name-disclosure leak rate---which is where coverage claims would matter. Obfuscation and encoding attacks are left to dedicated robustness benchmarks~\cite{debenedetti2024agentdojo,zhan2024injecagent}.

\textbf{Default visibility for undeclared components.} A tool, prompt, or resource registered without a \texttt{@require\_permissions} annotation carries no requirement and is therefore unrestricted by default: any authenticated caller sees it in listings and can invoke it, regardless of role or scope (\Cref{sec:component-filter,sec:require-permissions}). This is an intentional default---requiring an explicit annotation before a component is usable would slow ordinary development.

\textbf{Credential validation and revocation.} Cached verification may accept a revoked token for up to one TTL (\Cref{sec:adv-stale-cred}); deployments requiring immediate revocation must disable caching or add revocation signaling.

\textbf{Metadata exposure and evolution.} The pre-auth metadata route is optional and can be disabled or independently protected; its disclosure tradeoff is quantified in \Cref{sec:adv-registry-attacker}. Results apply to the versioned SDK snapshot in \Cref{sec:appendix-versions}. The MCP specification itself has also evolved since our original submission (2026-07-15): the Enterprise-Managed Authorization extension went stable 2026-06-18, and a broader specification revision followed 2026-07-28~\cite{mcp2026spec202607}; the sessionless revision (SEP-2567~\cite{mcp2026sep2567}) confirms that list endpoints may still vary by per-request authorization, consistent with how \texttt{PermissionFilterMiddleware} (C4) operates. As MCP and its SDKs evolve, these non-invasive layers can be retired where equivalent upstream capabilities emerge.

\section{Related Work}
\label{sec:related}

\textbf{Zero-trust and access control.} NIST SP~800-207~\cite{rose2020zerotrust} formalizes the zero-trust model, tracing to Saltzer-Schroeder least privilege~\cite{saltzer1975protection}; BeyondCorp~\cite{ward2014beyondcorp} and Microsoft's framework~\cite{microsoft2023zerotrust} demonstrate enterprise deployment at scale. We apply these ideas to AI-agent tool protocols. Our \texttt{@require\_permissions} mechanism is a lightweight declarative RBAC~\cite{sandhu1996rbac} variant in DNF, integrated into the tool-registration lifecycle so permissions drive discovery, filtering, and invocation enforcement from a shared annotation; unlike richer engines such as OPA~\cite{opa2024} or AWS Cedar~\cite{cedar2023}, or the ABAC~\cite{hu2014abac} and capability~\cite{dennis1966programming} models, it does not express resource-scoped, time-aware, or delegation-aware policies.

\textbf{OAuth and token frameworks.} RFC~7662~\cite{rfc7662} standardizes OAuth 2.0 introspection but requires client credentials at the introspection endpoint; \texttt{BaseTokenVerifier} extends this to vendor-specific authentication schemes and wraps the call in TTL caching and a scope-mapping hook. RFC~8693~\cite{rfc8693} (token exchange) is future work---our MCP server is a leaf in the call graph, not a delegator.

\textbf{AI-agent security.} Perez and Ribeiro~\cite{perez2022ignore} formalized the prompt-injection threat; Greshake et al.~\cite{greshake2023indirect} demonstrated indirect prompt injection through retrieved content; ToolEmu~\cite{ruan2023sandbox}, InjecAgent~\cite{zhan2024injecagent}, and AgentDojo~\cite{debenedetti2024agentdojo} provide systematic benchmarks for agent sandboxing and tool-call manipulation, and Liu et al.~\cite{liu2024promptinjection} formalize the attack/defense space. Input-side defenses such as StruQ~\cite{chen2024struq} and Spotlighting~\cite{hines2024spotlighting} reduce the probability that the model follows injected instructions, but the architectural defense we propose operates at a different layer: even if such defenses fail, the server-side authorization layer rejects the tool call.

\textbf{MCP security and ecosystem studies.} Work on MCP security spans several layers. At the gateway/centralized-governance layer, Matsumoto et al.~\cite{matsumoto2025a2arouting} propose a unified authentication and governance platform across A2A and MCP traffic; a companion system from our own deployment~\cite{kumar2026gateway} centralizes credential ingestion and persona resolution ahead of a server fleet (relationship detailed below); and MCP's own Enterprise-Managed Authorization extension~\cite{mcp2026ema} standardizes the same admission layer via ID-JAG token exchange. All three govern whether a caller reaches a server, not what it may discover or invoke once inside. At the ecosystem-measurement layer, Narajala and Habler~\cite{narajala2025enterprise}, Errico et al.~\cite{errico2025securing}, and Li and Gao~\cite{li2025mcpecosystem} survey enterprise MCP risks and registry-vetting weaknesses. At the host-side runtime layer, Xing et al.'s MCP-Guard~\cite{xing2025mcpguard} classifies malicious tool-call content in model output, not server-side authorization state. We are not aware of prior work that keeps a single server's pre-auth metadata, tool listing, and invocation decisions consistent with each other, which is what this paper does.

\textbf{FastMCP-native and policy-engine authorization.} FastMCP's own \texttt{auth=} parameter~\cite{fastmcp2026authorization} and Cerbos's FastMCP integration~\cite{cerbos2025fastmcp} both couple tool listing and invocation to policy, closing part of the gap in \Cref{sec:gap-analysis}---FastMCP via a single conjunctive callable per component, Cerbos via separate \texttt{tools/list::}/\texttt{tools/call::} actions at an external decision point. \Cref{sec:require-permissions} details what we add on top of both: disjunctive (DNF) composition and a third consistent surface, pre-auth metadata (C3), derived from the same declaration rather than maintained separately.

\textbf{Relationship to our enterprise MCP gateway.} The companion system cited above~\cite{kumar2026gateway} centralizes credential ingestion and persona resolution \emph{upstream} of any individual MCP server: it decides whether a caller reaches a server, not what an authenticated caller may discover or invoke once inside, which is what this paper addresses. Our server independently verifies issuer, expiry, and scope regardless of upstream admission (\Cref{sec:threat-model}); \citeauthor{kumar2026gateway} describe this same split themselves, calling server-side authorization ``a concurrent line'' distinct from their gateway. The split applies equally to MCP's own Enterprise-Managed Authorization extension~\cite{mcp2026ema}.

\textbf{Positioning against existing stacks.} \Cref{tab:positioning} summarizes the comparison; every row leaves the MCP protocol unmodified. Service meshes and API gateways~\cite{li2019servicemesh}, the gateway system above~\cite{kumar2026gateway}, and MCP's own EMA extension~\cite{mcp2026ema} terminate authentication at the HTTP/fleet boundary, not inside a server's discovery and invocation surfaces. Policy engines (OPA~\cite{opa2024}, Cedar~\cite{cedar2023}) express richer, delegation-aware policies but are not MCP-native and do not couple to pre-auth metadata---future work behind our decorator (\Cref{sec:conclusion}). SPIFFE/SPIRE~\cite{spiffe2022} solves workload identity, not per-tool authorization. We are not aware of a prior system combining all four columns of \Cref{tab:positioning} in one MCP-native package deployed in production (\Cref{sec:perf}), though listing-plus-invocation coupling itself is now available natively in FastMCP.

\begin{table}[htbp]
\centering
\caption{Positioning vs.\ adjacent stacks. ``n/r'' = not reported.}
\label{tab:positioning}
\footnotesize
\begin{tabular*}{\columnwidth}{@{\extracolsep{\fill}}lcccc@{}}
\toprule
                              & \rotatebox{75}{Pre-auth schema} & \rotatebox{75}{Per-tool authz} & \rotatebox{75}{Cross-header creds} & \rotatebox{75}{Discovery-coupled} \\
\midrule
Envoy / Istio~\cite{li2019servicemesh}     & --- & --- & part. & --- \\
OPA / Cedar~\cite{opa2024,cedar2023}       & --- & yes & ---   & --- \\
SPIFFE/SPIRE~\cite{spiffe2022}             & --- & --- & ---   & --- \\
FastMCP \texttt{auth=}~\cite{fastmcp2026authorization} & --- & yes & part. & part. \\
Cerbos--FastMCP~\cite{cerbos2025fastmcp}   & --- & yes & ---   & part. \\
Gateway~\cite{kumar2026gateway}            & n/r & n/r & yes   & n/r \\
EMA (MCP ext.)~\cite{mcp2026ema}           & --- & --- & ---   & --- \\
\textbf{This work}                          & \textbf{yes} & \textbf{yes} & \textbf{yes} & \textbf{yes} \\
\bottomrule
\end{tabular*}
\end{table}

\section{Conclusion and Future Work}
\label{sec:conclusion}

We presented composable extensions (cross-header credential normalization, pluggable token verification with caching, pre-authentication discovery, and permission-aware visibility kept consistent with per-tool invocation enforcement) that bring NIST SP~800-207 zero-trust principles to enterprise MCP deployments without modifying the protocol or the core SDK. A single shared permission annotation drives discovery, listing, and per-tool enforcement from MCP's structured metadata, which opaque HTTP endpoints do not expose. Several directions remain:

\begin{itemize}
  \item \textbf{Policy-as-code integration.} Slotting OPA or Cedar in behind \texttt{@require\_permissions} for time-based, resource-scoped, or delegation-aware policies.
  \item \textbf{stdio security.} A capability-based scheme or mutual TLS over local sockets for the stdio transport, which currently bypasses HTTP security entirely.
  \item \textbf{Token exchange (RFC~8693).} Maintaining the zero-trust chain when an MCP server calls a downstream service on a user's behalf---the \emph{outbound} direction, distinct from EMA's \emph{inbound} ID-JAG admission~\cite{mcp2026ema}.
\end{itemize}

\bibliographystyle{ACM-Reference-Format}
\bibliography{references}

@misc{anthropic2024mcp,
  author       = {{Anthropic}},
  title        = {Model Context Protocol Specification},
  howpublished = {\url{https://modelcontextprotocol.io}},
  year         = {2024}
}

@techreport{rose2020zerotrust,
  author      = {Scott Rose and Oliver Borchert and Stu Mitchell and Sean Connelly},
  title       = {Zero Trust Architecture},
  institution = {National Institute of Standards and Technology},
  type        = {Special Publication},
  number      = {800-207},
  year        = {2020}
}

@misc{rfc8414,
  author       = {Michael Jones and Nat Sakimura and John Bradley},
  title        = {{OAuth} 2.0 Authorization Server Metadata},
  howpublished = {RFC 8414, IETF},
  year         = {2018}
}

@misc{rfc9728,
  author       = {Michael Jones and Phil Hunt and Aaron Parecki},
  title        = {{OAuth} 2.0 Protected Resource Metadata},
  howpublished = {RFC 9728, IETF},
  year         = {2025}
}

@article{ward2014beyondcorp,
  author  = {Rory Ward and Betsy Beyer},
  title   = {{BeyondCorp}: A New Approach to Enterprise Security},
  journal = {;login:},
  volume  = {39},
  number  = {6},
  year    = {2014},
  publisher = {USENIX}
}

@misc{microsoft2023zerotrust,
  author       = {{Microsoft}},
  title        = {Zero Trust deployment guide},
  howpublished = {Microsoft Learn, \url{https://learn.microsoft.com/en-us/security/zero-trust/}},
  year         = {2023}
}

@inproceedings{li2019servicemesh,
  author    = {Wubin Li and Yves Lemieux and Jing Gao and Zhuofeng Zhao and Yanbo Han},
  title     = {Service Mesh: Challenges, State of the Art, and Future Research Opportunities},
  booktitle = {IEEE International Conference on Service-Oriented System Engineering},
  year      = {2019}
}

@misc{rfc7662,
  author       = {Justin Richer},
  title        = {{OAuth} 2.0 Token Introspection},
  howpublished = {RFC 7662, IETF},
  year         = {2015}
}

@misc{opa2024,
  author       = {{Open Policy Agent}},
  title        = {Policy-based control for cloud native environments},
  howpublished = {\url{https://www.openpolicyagent.org}},
  year         = {2024}
}

@misc{cedar2023,
  author       = {{Amazon Web Services}},
  title        = {Cedar: A language for defining permissions as policies},
  howpublished = {\url{https://www.cedarpolicy.com}},
  year         = {2023}
}

@inproceedings{ruan2023sandbox,
  author    = {Yangjun Ruan and Honghua Dong and Andrew Wang and Silviu Pitis and Yongchao Zhou and Jimmy Ba and Yann Dubois and Chris J. Maddison and Tatsunori Hashimoto},
  title     = {Identifying the Risks of {LM} Agents with an {LM}-Emulated Sandbox},
  booktitle = {International Conference on Learning Representations (ICLR)},
  year      = {2024}
}

@inproceedings{zhan2024injecagent,
  author    = {Qiusi Zhan and Zhixiang Liang and Zifan Ying and Daniel Kang},
  title     = {{InjecAgent}: Benchmarking Indirect Prompt Injections in Tool-Integrated Large Language Model Agents},
  booktitle = {Findings of the Association for Computational Linguistics (ACL Findings)},
  year      = {2024}
}

@article{saltzer1975protection,
  author  = {Jerome H. Saltzer and Michael D. Schroeder},
  title   = {The Protection of Information in Computer Systems},
  journal = {Proceedings of the IEEE},
  volume  = {63},
  number  = {9},
  pages   = {1278--1308},
  year    = {1975}
}

@article{sandhu1996rbac,
  author  = {Ravi S. Sandhu and Edward J. Coyne and Hal L. Feinstein and Charles E. Youman},
  title   = {Role-Based Access Control Models},
  journal = {IEEE Computer},
  volume  = {29},
  number  = {2},
  pages   = {38--47},
  year    = {1996}
}

@techreport{hu2014abac,
  author      = {Vincent C. Hu and David Ferraiolo and Rick Kuhn and Adam Schnitzer and Kenneth Sandlin and Robert Miller and Karen Scarfone},
  title       = {Guide to Attribute Based Access Control ({ABAC}) Definition and Considerations},
  institution = {National Institute of Standards and Technology},
  type        = {Special Publication},
  number      = {800-162},
  year        = {2014}
}

@article{dennis1966programming,
  author  = {Jack B. Dennis and Earl C. Van Horn},
  title   = {Programming Semantics for Multiprogrammed Computations},
  journal = {Communications of the ACM},
  volume  = {9},
  number  = {3},
  pages   = {143--155},
  year    = {1966}
}

@misc{rfc8693,
  author       = {Michael Jones and Anthony Nadalin and Brian Campbell and John Bradley and Chuck Mortimore},
  title        = {{OAuth} 2.0 Token Exchange},
  howpublished = {RFC 8693, IETF},
  year         = {2020}
}

@inproceedings{greshake2023indirect,
  author    = {Kai Greshake and Sahar Abdelnabi and Shailesh Mishra and Christoph Endres and Thorsten Holz and Mario Fritz},
  title     = {Not what you've signed up for: {C}ompromising Real-World {LLM}-Integrated Applications with Indirect Prompt Injection},
  booktitle = {Proceedings of the 16th ACM Workshop on Artificial Intelligence and Security (AISec)},
  year      = {2023}
}

@misc{acm-policy-ai,
  author       = {{Association for Computing Machinery}},
  title        = {{ACM} Policy on Authorship},
  howpublished = {\url{https://www.acm.org/publications/policies/new-acm-policy-on-authorship}},
  year         = {2023}
}

@inproceedings{debenedetti2024agentdojo,
  author    = {Edoardo Debenedetti and Jie Zhang and Mislav Balunovi\'c and Luca Beurer-Kellner and Marc Fischer and Florian Tram\`er},
  title     = {{AgentDojo}: A Dynamic Environment to Evaluate Prompt Injection Attacks and Defenses for {LLM} Agents},
  booktitle = {Advances in Neural Information Processing Systems (NeurIPS)},
  year      = {2024}
}

@inproceedings{liu2024promptinjection,
  author    = {Yupei Liu and Yuqi Jia and Runpeng Geng and Jinyuan Jia and Neil Zhenqiang Gong},
  title     = {Formalizing and Benchmarking Prompt Injection Attacks and Defenses},
  booktitle = {33rd USENIX Security Symposium},
  year      = {2024}
}

@misc{chen2024struq,
  author       = {Sizhe Chen and Julien Piet and Chawin Sitawarin and David Wagner},
  title        = {{StruQ}: Defending Against Prompt Injection with Structured Queries},
  howpublished = {arXiv preprint arXiv:2402.06363},
  year         = {2024}
}

@misc{hines2024spotlighting,
  author       = {Keegan Hines and Gary Lopez and Matthew Hall and Federico Zarfati and Yonatan Zunger and Emre Kiciman},
  title        = {Defending Against Indirect Prompt Injection Attacks With Spotlighting},
  howpublished = {arXiv preprint arXiv:2403.14720},
  year         = {2024}
}

@misc{owasp2025llmtop10,
  author       = {{OWASP Foundation}},
  title        = {{OWASP} Top 10 for Large Language Model Applications, 2025 Edition},
  howpublished = {\url{https://owasp.org/www-project-top-10-for-large-language-model-applications/}},
  year         = {2025}
}

@inproceedings{hardy1988confused,
  author    = {Norm Hardy},
  title     = {The Confused Deputy: (or why capabilities might have been invented)},
  booktitle = {ACM SIGOPS Operating Systems Review},
  volume    = {22},
  number    = {4},
  pages     = {36--38},
  year      = {1988}
}

@article{wilson1927ci,
  author  = {Edwin B. Wilson},
  title   = {Probable Inference, the Law of Succession, and Statistical Inference},
  journal = {Journal of the American Statistical Association},
  volume  = {22},
  number  = {158},
  pages   = {209--212},
  year    = {1927}
}

@techreport{nist2023airmf,
  author      = {{National Institute of Standards and Technology}},
  title       = {Artificial Intelligence Risk Management Framework ({AI RMF} 1.0)},
  institution = {NIST},
  number      = {NIST AI 100-1},
  year        = {2023}
}

@misc{spiffe2022,
  author       = {{SPIFFE Project / CNCF}},
  title        = {{SPIFFE}: Secure Production Identity Framework for Everyone},
  howpublished = {\url{https://spiffe.io/docs/latest/spiffe-about/spiffe-concepts/}},
  year         = {2022}
}

@inproceedings{perez2022ignore,
  author    = {F\'abio Perez and Ian Ribeiro},
  title     = {Ignore Previous Prompt: Attack Techniques For Language Models},
  booktitle = {NeurIPS ML Safety Workshop},
  year      = {2022}
}

@inproceedings{narajala2025enterprise,
  author    = {Vineeth Sai Narajala and Idan Habler},
  title     = {Enterprise-Grade Security for the Model Context Protocol ({MCP}): Frameworks and Mitigation Strategies},
  booktitle = {2026 IEEE 5th International Conference on AI in Cybersecurity (ICAIC)},
  year      = {2026},
  note      = {Originally circulated as arXiv:2504.08623 (2025)}
}

@inproceedings{matsumoto2025a2arouting,
  author    = {Daichi Matsumoto and Koki Watarai and Satoshi Okada and Takuho Mitsunaga},
  title     = {A2A Routing Service with {MCP} Integration: Bridging Security, Auditability, and Governance in {AI} Agent Systems},
  booktitle = {2025 IEEE International Conference on Computing (ICOCO)},
  year      = {2025}
}

@misc{kumar2026gateway,
  author       = {Suraj Kumar and Amy Wang and Srinivasan Manoharan},
  title        = {A Gateway Architecture for Enterprise {MCP} Authentication: Unifying Heterogeneous Auth, Identity Delegation, and the User / Non-User Persona Problem},
  howpublished = {arXiv preprint arXiv:2608.10760},
  year         = {2026}
}

@inproceedings{xing2025mcpguard,
  author    = {Wenpeng Xing and Zhonghao Qi and Yupeng Qin and Yilin Li and Caini Chang and Jiahui Yu and Changting Lin and Zhenzhen Xie and Meng Han},
  title     = {{MCP-Guard}: A Multi-Stage Defense-in-Depth Framework for Securing Model Context Protocol in Agentic {AI}},
  booktitle = {Findings of the Association for Computational Linguistics: ACL 2026},
  pages     = {4877--4889},
  year      = {2026},
  publisher = {Association for Computational Linguistics},
  doi       = {10.18653/v1/2026.findings-acl.240},
  note      = {Originally circulated as arXiv:2508.10991 (2025)}
}

@misc{fastmcp2026authorization,
  author       = {Jeremiah Lowin},
  title        = {{FastMCP} Authorization: Component-Level and Server-Level Access Control},
  howpublished = {\url{https://gofastmcp.com/servers/authorization}},
  year         = {2026}
}

@misc{cerbos2025fastmcp,
  author       = {{Cerbos}},
  title        = {How to Secure Your {FastMCP} Server With Permission Management},
  howpublished = {\url{https://www.cerbos.dev/blog/how-to-secure-your-fast-mcp-server-with-permission-management}},
  year         = {2025}
}

@misc{errico2025securing,
  author       = {Herman Errico and Jiquan Ngiam and Shanita Sojan},
  title        = {Securing the Model Context Protocol ({MCP}): Risks, Controls, and Governance},
  howpublished = {arXiv preprint arXiv:2511.20920},
  year         = {2025}
}

@inproceedings{li2025mcpecosystem,
  author    = {Xiaofan Li and Xing Gao},
  title     = {A First Look at the Security Issues in the Model Context Protocol Ecosystem},
  booktitle = {Proceedings of the 56th Annual IEEE/IFIP International Conference on Dependable Systems and Networks (DSN)},
  year      = {2026},
  note      = {Originally circulated as arXiv:2510.16558}
}

@misc{mcp2026ema,
  author       = {{Anthropic}},
  title        = {{MCP} Extension: Enterprise-Managed Authorization},
  howpublished = {\url{https://modelcontextprotocol.io/extensions/auth/enterprise-managed-authorization}},
  year         = {2026}
}

@misc{mcp2026spec202607,
  author       = {{Anthropic}},
  title        = {The Model Context Protocol Specification (2026-07-28)},
  howpublished = {\url{https://modelcontextprotocol.io/specification/2026-07-28/changelog}},
  year         = {2026}
}

@misc{mcp2026sep2567,
  author       = {Peter Alexander},
  title        = {{SEP-2567}: Sessionless {MCP} via Explicit State Handles},
  howpublished = {\url{https://modelcontextprotocol.io/seps/2567-sessionless-mcp}},
  year         = {2026},
  note         = {Model Context Protocol Enhancement Proposal, Final}
}

\appendix

\section{MCP SDK Versions Analyzed}
\label{sec:appendix-versions}

The cross-SDK gap analysis in \Cref{sec:gap-analysis} is based on the SDK versions listed in \Cref{tab:sdk-versions}, reflecting the latest stable releases as of 2026-05-15. These six are a sample of the ten official SDKs; Java, Kotlin, PHP, and Ruby are outside our empirical claims. Later releases may close some gaps; the contributions remain valid as a transitional reference architecture and a portability target for upstream adoption (\Cref{sec:discussion}).
\begin{table}[h!]
\centering
\caption{MCP SDK versions analyzed.}
\label{tab:sdk-versions}
\footnotesize
\begin{tabular*}{\columnwidth}{@{\extracolsep{\fill}}lll@{}}
\toprule
Language & Version & Released \\
\midrule
Python     & v1.27.0 (+ \texttt{fastmcp} v3.4.0) & 2026 \\
TypeScript & v1.26.0                            & 2026-02-04 \\
Go         & v1.3.0                             & 2026-02-09 \\
Rust       & \texttt{rmcp} v1.2.0               & 2026-03-11 \\
C\#        & v1.1.0                             & 2026 \\
Swift      & 0.10.2                             & 2025-09-23 \\
\bottomrule
\end{tabular*}
\end{table}

All repositories are under \texttt{modelcontextprotocol/}. Python is the subject of our extensions; for TypeScript we analyzed the v1.x production line (v2 was pre-alpha on \texttt{main}); the C\# SDK is the latest stable tag following v1.0 (still marked ``preview'').

\section*{Open Science}
\label{sec:open-science}

The artifacts supporting this submission will be made available in a public repository (link to be added at submission/acceptance): the FastMCP extension code implementing C1--C4, the eight-tool mock MCP server harness for the three evaluated configurations (baseline / filter-only / with-extensions), the 60 hand-reviewed prompt-injection payloads, the multi-model sweep driver with its raw 2160-attempt outputs and aggregation scripts, the scripted name-guess attacker (\Cref{sec:adv-name-guess}), and the dual-persona and token-verifier deployment configurations (\Cref{sec:multi-auth}, \Cref{sec:token-verifier}). We do not release the gateway-side LLM client: \texttt{gpt-5}, \texttt{claude-sonnet-4-6}, \texttt{claude-opus-4-7}, and \texttt{gemini-2.5-pro} are accessed through a proprietary enterprise gateway---but the harness speaks an OpenAI-compatible chat-completion API and re-runs against any OpenAI-compatible endpoint serving these models. Production metrics in \Cref{sec:perf} are observational and not reproducible from the artifact. 

\section*{Ethical Considerations}
\label{sec:ethics}

The work described in this paper is a defensive contribution: we propose architecture and code to prevent unauthorized calls to privileged tools by LLM agents. The adversarial component of our evaluation (\Cref{sec:adv-prompt-injection}, 2160 attempts across four LLMs) uses a synthetic, LLM-drafted and hand-reviewed corpus of 60 prompt-injection payloads against mock MCP tools (\texttt{restart\_pod} etc., implemented as no-op stubs); no real system, user, or third-party provider is harmed or load-tested in the process. No human subjects are involved. The prompt corpus is released alongside the code (see \emph{Open Science}); we view its release as net-positive: similar payloads can already be elicited from off-the-shelf jailbreak repositories, and publishing them in a controlled benchmark form serves defenders more than it informs attackers. The vulnerabilities we surface in the MCP SDK design (\Cref{sec:gap-analysis}) are properties of the documented public API rather than secret defects, and the SDK maintainers' public issue trackers already contain related discussions; no responsible-disclosure pathway was bypassed. Production-deployment statistics in \Cref{sec:perf} are aggregate and contain no per-user data.

\section*{Declaration of Generative AI and AI-assisted technologies in the writing process}
\label{sec:genai-declaration}

The generative-AI tools used in making this paper were Anthropic Claude (Opus 4 family and Sonnet 5, via the Anthropic API and Claude Code) and OpenAI ChatGPT (GPT-5 family, web and API). They were used in five scoped ways, with the authors reviewing, revising, and testing every retained output: (i)~assisting with drafting, editing, and formatting the paper's prose; (ii)~drafting the prompt-injection corpus (\Cref{sec:adv-prompt-injection}) against a documented per-class hook list, every prompt hand-reviewed, following InjecAgent~\cite{zhan2024injecagent}; (iii)~drafting boilerplate harness code, with all attack and aggregation logic verified by the authors; (iv)~coding assistance for the authors' implementation of the extension code (Claude Code); the released open-source SDK was then derived from that code with AI assistance to strip internal references; and (v)~surfacing candidate references, each of which the authors verified against its source (arXiv ID, DOI, ACL Anthology, IEEE Xplore, or publisher page) for title, authors, year, and venue---discarding any that could not be cross-checked, with an automated re-verification pass before submission. These same models are also subjects of the evaluation in \Cref{sec:adv-prompt-injection}, as disclosed there. AI tools were \emph{not} used to fabricate results, invent citations, or substitute for authorial reasoning; use complied with the ACM Policy on Authorship~\cite{acm-policy-ai}, and the authors take full responsibility for the paper.

\end{document}